

\input epsf.tex

\font\rmu=cmr10 scaled\magstephalf
\font\bfu=cmbx10 scaled\magstephalf

\font\it=cmti10 scaled \magstephalf

\rmu

\font\rmus=cmr8
\font\rmuss=cmr6
\font\mait=cmmi10 scaled\magstephalf
\font\maits=cmmi7 scaled\magstephalf
\font\maitss=cmmi7
\font\msyb=cmsy10 scaled\magstephalf
\font\msybs=cmsy8 scaled\magstephalf
\font\msybss=cmsy7
\font\bfus=cmbx7 scaled\magstephalf
\font\bfuss=cmbx7
\font\cmeq=cmex10 scaled\magstephalf

\textfont0=\rmu
\scriptfont0=\rmus
\scriptscriptfont0=\rmuss

\textfont1=\mait
\scriptfont1=\maits
\scriptscriptfont1=\maitss

\textfont2=\msyb
\scriptfont2=\msybs
\scriptscriptfont2=\msybss

\textfont3=\cmeq
\scriptfont3=\cmeq
\scriptscriptfont3=\cmeq

\newfam\bmufam  \textfont\bmufam=\bfu
      \scriptfont\bmufam=\bfus \scriptscriptfont\bmufam=\bfuss

\hsize=15.5cm
\vsize=21cm
\baselineskip=16pt   
\parskip=12pt plus  2pt minus 2pt

\def\a{\alpha}
\def\b{\beta}
\def\d{\delta}
\def\e{\epsilon}

\def\g{\gamma}

\def\semi{\bigcirc\kern-1em{s}\;}

\def\del{\partial}
\def\ni{\noindent}
\def\R{{\rm I\!R}}

\def\one{{\mathchoice {\rm 1\mskip-4mu l} {\rm 1\mskip-4mu l}
{\rm 1\mskip-4.5mu l} {\rm 1\mskip-5mu l}}}
\def\Q{{\mathchoice
{\setbox0=\hbox{$\displaystyle\rm Q$}\hbox{\raise 0.15\ht0\hbox to0pt
{\kern0.4\wd0\vrule height0.8\ht0\hss}\box0}}
{\setbox0=\hbox{$\textstyle\rm Q$}\hbox{\raise 0.15\ht0\hbox to0pt
{\kern0.4\wd0\vrule height0.8\ht0\hss}\box0}}
{\setbox0=\hbox{$\scriptstyle\rm Q$}\hbox{\raise 0.15\ht0\hbox to0pt
{\kern0.4\wd0\vrule height0.7\ht0\hss}\box0}}
{\setbox0=\hbox{$\scriptscriptstyle\rm Q$}\hbox{\raise 0.15\ht0\hbox to0pt
{\kern0.4\wd0\vrule height0.7\ht0\hss}\box0}}}}
\def\C{{\mathchoice
{\setbox0=\hbox{$\displaystyle\rm C$}\hbox{\hbox to0pt
{\kern0.4\wd0\vrule height0.9\ht0\hss}\box0}}
{\setbox0=\hbox{$\textstyle\rm C$}\hbox{\hbox to0pt
{\kern0.4\wd0\vrule height0.9\ht0\hss}\box0}}
{\setbox0=\hbox{$\scriptstyle\rm C$}\hbox{\hbox to0pt
{\kern0.4\wd0\vrule height0.9\ht0\hss}\box0}}
{\setbox0=\hbox{$\scriptscriptstyle\rm C$}\hbox{\hbox to0pt
{\kern0.4\wd0\vrule height0.9\ht0\hss}\box0}}}}

\font\fivesans=cmss10 at 4.61pt
\font\sevensans=cmss10 at 6.81pt
\font\tensans=cmss10
\newfam\sansfam
\textfont\sansfam=\tensans\scriptfont\sansfam=\sevensans\scriptscriptfont
\sansfam=\fivesans
\def\sans{\fam\sansfam\tensans}
\def\Z{{\mathchoice
{\hbox{$\sans\textstyle Z\kern-0.4em Z$}}
{\hbox{$\sans\textstyle Z\kern-0.4em Z$}}
{\hbox{$\sans\scriptstyle Z\kern-0.3em Z$}}
{\hbox{$\sans\scriptscriptstyle Z\kern-0.2em Z$}}}}

\newcount\foot
\foot=1
\def\note#1{\footnote{${}^{\number\foot}$}{\ftn #1}\advance\foot by 1}

\def\frac#1#2{{#1\over #2}}
\def\text#1{\quad{\hbox{#1}}\quad}

\font\ch=cmbx12 scaled\magstephalf
\font\ftn=cmr8 scaled\magstephalf

\font\it=cmti10 scaled\magstephalf

\font\titch=cmbx12 scaled\magstep2
\font\titname=cmr10 scaled\magstep2
\font\titit=cmti10 scaled\magstep1
\font\titbf=cmbx10 scaled\magstep2

\nopagenumbers

\line{\hfil DFF 220/01/95}
\line{\hfil Jan 31, 1995}
\vskip2.2cm
\centerline{\titch NON-PERTURBATIVE SOLUTIONS FOR}
\vskip.5cm
\centerline{\titch LATTICE QUANTUM GRAVITY}
\vskip1.7cm
\centerline{\titname R. Loll\note{Supported by the European Human
Capital and Mobility program on ``Constrained Dynamical Systems"}}
\vskip.5cm
\centerline{\titit Sezione INFN di Firenze}
\vskip.2cm
\centerline{\titit Largo E. Fermi 2}
\vskip.2cm
\centerline{\titit I-50125 Firenze, Italy}

\vskip2.2cm
\centerline{\titbf Abstract}

We propose a new, discretized model for the study of 3+1-dimensional
canonical quantum gravity, based on the classical
$SL(2,\C)$-connection formulation. The discretization takes place on
a topological $N^3$-lattice with periodic boundary conditions.
All operators and wave functions are constructed from
one-dimensional link variables, which are regarded as the fundamental
building blocks of the theory. The kinematical Hilbert space is
spanned by polynomials of certain
Wilson loops on the lattice and is manifestly
gauge- and diffeomorphism-invariant. The discretized quantum
Hamiltonian $\hat H$ maps this space into itself. We find a large
sector of solutions to the discretized Wheeler-DeWitt equation
$\hat H\psi=0$, which are labelled by single and multiple Polyakov
loops. These states have a finite norm with respect to a natural
scalar product on the space of holomorphic $SL(2,\C)$-Wilson loops.
We also investigate the existence of further solutions for the case of
the $1^3$-lattice. - Our results provide for the first time a
rigorous, regularized framework for studying non-perturbative quantum
gravity.

\vfill\eject
\footline={\hss\tenrm\folio\hss}
\pageno=1


\line{\ch 1 Introduction\hfil}

All attempts to define a lattice discretization of four-dimensional
quantum gravity, in analogy with the rather powerful methods employed
in quantum chromodynamics, have been riddled with difficulties.
A major problem is that of incorporating the
diffeomorphism-invariance of the theory. The discretization typically
destroys this symmetry already at the classical level, similar to the
way in which a lattice  discretization
of Yang-Mills theory breaks its rotational symmetry. However, in the
case of gravity this is much more serious, since the diffeomorphism
group is infinite-dimensional, and acts non-linearly on the underlying
manifold. A number of
questions arise: (i) does the discretized quantum
theory possess a residual diffeomorphism symmetry? (ii) how is the
continuum limit to be taken? and (iii) what does the resulting theory
look like? (This should be a non-perturbative,
diffeomorphism-invariant description for quantum gravity in an
``unbroken" phase.) The
answer to (i) is usually ``no", whereas the other two questions are
hard to address and have not yet found a satisfactory solution within
either Lagrangian or Hamiltonian approaches. Using
Euclidean path integral methods, a main problem is that of  finding
the correct measure, without over- or undercounting the physical
configurations (currently the most active research programs are
quantum Regge calculus and dynamically
triangulated gravity; see [1] for a recent review). A central
problem in the canonical approaches is that of finding a closing
algebra of  diffeomorphism constraints on the phase space of the
regularized theory (see, for example, the discussion in [2]). The
unresolved status of the closure problem is the reason for which many
Hamiltonian discretized models have not advanced very far [3,2].

The starting point of this paper is a particular form of
Hamiltonian lattice gravity. However, instead of
implementing the action of the (three-dimensional) spatial
diffeomorphisms directly on the phase space of the theory, we will use
a framework in which diffeomorphism invariance is manifest.
This approach is inspired by the
loop space formulation of canonical quantum gravity, based on the
classical reformulation of Einstein gravity in terms of a pair $(A_a^i,
\tilde E_i^a)$ of
$SL(2,\C)$-Yang-Mills variables due to Ashtekar [4,5]. In the
original paper by Rovelli and Smolin [6], the wave functions
$\psi(\g)$ of the quantum theory are labelled by spatial loops $\g$,
and the diffeomorphism invariance is formally implemented by
selecting those wave functions that are invariant when the loop
argument $\g$ is moved by a diffeomorphism. Physical wave functions
therefore depend on diffeomorphism equivalence classes $[\g]$ of
closed loops. Using these ideas, solutions to all of the quantum
constraints, including the Wheeler-DeWitt equation $\hat H\psi=0$, have
been found [6] (for an overview of this and other solutions, see [7]).
This involves the choice of a regularization and factor-ordering for
the quantum Hamiltonian constraint $\hat H$, and is rather formal, in
the sense that there is no well-defined scalar product  on the space
of solutions, and little control over the influence of
different regularizations and factor orderings on the structure of
this space. As a result the status of these solutions has remained
unclear and the issue of reality conditions could not be
addressed. In the Ashtekar formulation in terms of
complex canonical variable pairs $(A_a^i,\tilde E_i^a)$, such
conditions have to be implemented in the quantum theory to make sure
that real, and not complex gravity is described.

Recently, there have been proposals for defining continuum loop
representations rigorously , regarding them
as non-linear analogues of the
Fock representation based on quantum loop states (see [8] for a review
and further references). Central to this line of research are
the construction of suitable domains for wave functions depending on
connections modulo gauge transformations and diffeomorphism-invariant
measures on such spaces. However, so far these efforts have addressed
only the kinematical structure of the quantum theory, without
incorporating any of the dynamical issues regarding the quantum
Hamiltonian. A rigorous analysis of the structure of the solution
space to the Wheeler-DeWitt equation and of physical operators on this
space is therefore still lacking. In the present work we will suggest
a discretized version of canonical quantum gravity that can deal with
these issues.

The discretization will take place on a periodic cubic lattice,
but this is not to be thought of as a fixed lattice embedded in
physical, Euclidean three-space, but merely as a topological quantity,
defined by those of its properties that would remain unaffected by a
smooth diffeomorphism. It may thus be thought of as representing a
diffeomorphism equivalence class of cubic lattices. All physical
quantities are defined in terms of link variables, and the discretized
Hamiltonian  acts purely combinatorially on wave functions labelled
by  lattice loops.  The formalism is manifestly diffeomorphism- and
therefore also scale-invariant. The basic link variables of our
formulation are a Kogut-Susskind pair of an $SL(2,\C)$-valued link
holonomy and a corresponding $sl(2,\C)$-momentum variable. However,
the corresponding basic operators in the quantum representation are
{\it not} self-adjoint. This is acceptable, since they do not
correspond to any physical observables. It comes about since we
propose to implement the reality conditions, following the ideas of
[5,3], as a holomorphicity condition on wave functions (this leads to
the correct number of degrees of freedom in the quantum theory). Also
the Hamiltonian operator turns out to be non-selfadjoint. This is no
real reason for concern either, if one accepts the argument that the
self-adjointness of the Hamiltonian in a generally covariant theory is
not a strict physical requirement (for related discussions, see [9]).

On a cubic $N^3$-lattice with periodic boundary conditions, given a
specific discretization of the Hamiltonian $H$ together with a
particular operator ordering for $\hat H$, we are able to find an
infinite set of solutions to the Wheeler-DeWitt equation, that
moreover have finite norm with respect to the inner product induced
from the original kinematical
Hilbert space. They are labelled by so-called
Polyakov loops and multiple Polyakov loops, which are well-known from
their role as order parameters for the phase structure of lattice gauge
theory (see, for example, [10]). Their appearance in the context of
lattice gravity is curious, although they play here a quite
different role, namely, that of parametrizing the solution space to the
Wheeler-DeWitt equation. These solutions are the lattice analogues of
the non-intersecting, smooth-loop solutions found by Rovelli and Smolin
[6]. This for the first time provides a regularized model for
non-perturbative quantum gravity that can be used to study
physical observables.

The paper is organized as follows. In the next section we introduce
all the necessary ingredients for setting up the Hamiltonian lattice
theory, and explain some features of the holomorphic representation
we will be using for constructing a scalar product on $SL(2,\C)$-wave
functions. We derive an important explicit formula for relating
arbitrary polynomials in the $SU(2)$- and $SL(2,\C)$-theories.
In section 3, we establish a relation between the  zero-eigenvalue
solutions of our quantum Hamiltonian and another Hamiltonian, induced
from the $SU(2)$-theory. This suggests a close connection between the
real and the complex, holomorphic theories. Next we demonstrate that
there is a large subspace of the Hilbert space that is annihilated by
the Hamiltonian constraint. In section 4, we take a closer look at the
special case of the $1^3$-lattice and explore the possibility of
finding solutions beyond the ones labelled by the Polyakov loops. We
illustrate some of the technicalities involved, without being able to
identify any solutions explicitly. The last section contains our
conclusions and an outlook.

\vskip2cm

\line{\ch 2 The general formalism\hfil}

We first recall the basic Hamiltonian variables for the $SU(2)$-lattice
gauge theory, before discussing the complexified framework for
$SL(2,\C)=SU(2)_\C$. This is necessary for setting up a discretized
version of the connection formulation of canonical gravity. - With
each lattice link we associate an element $g\in SU(2)$, parametrized
by a matrix $V(g)$ in the defining two-dimensional representation as

$$
V_A{}^B=\left( \matrix{\a_0+i\a_1&\a_2+i\a_3\cr
    -\a_2+i\a_3&\a_0-i\a_1}\right),\eqno(2.1)
$$

\ni with $\a_i\in\R$, and subject to the condition $\sum_{i=0}^3
\a_i^2=1$. The matrix $V_A{}^B$ can be written as a (real) linear
combination of the unit matrix $\one$ and the three $\tau$-matrices
defined by

$$
\tau_1=\left(\matrix{i&0\cr 0&-i}\right),\quad
\tau_2=\left(\matrix{0&1\cr -1&0}\right),\quad
\tau_3=\left(\matrix{0&i\cr i&0}\right).\eqno(2.2)
$$

\ni The $\tau$-matrices satisfy $[\tau_i,\tau_j]=2\,
\e_{ijk}\tau_k$. The differential operator

$$
(\frac{\del}{\del V})_A{}^B=\frac14 \left(\matrix{
\del_0-i\del_1&-\del_2-i\del_3\cr\del_2-i\del_3&\del_0+i\del_1}
\right)\eqno(2.3)
$$

\ni acts on the representation matrices (2.1) and satisfies

$$
(\frac{\del}{\del V})_A{}^B\, V_C{}^D=\frac12\,\d_A{}^D\,
\d_C{}^B.\eqno(2.4)
$$

\ni The operators corresponding to the classical momentum variables
$p_i$ (with a gauge algebra index $i$) are given by

$$
\hat p_i=-i\,\tau_{iA}{}^B  V_B{}^C(\frac{\del}{\del
V})_C{}^A\eqno(2.5) $$

\ni and satisfy

$$
\eqalign{
&\hat p_i\, V_A{}^C=-\frac{i}{2}\,\tau_{iA}{}^B V_B{}^C\cr
&\hat p_i\, (V^{-1})_A{}^C=\frac{i}{2}(V^{-1})_A{}^B\tau_{iB}{}^C.}
\eqno(2.6)
$$

\ni In terms of coordinates, we have

$$
\eqalign{
&\hat p_1=\frac{i}{2} (\a_1\del_0-\a_0\del_1+\a_3\del_2-\a_2\del_3)\cr
&\hat p_2=\frac{i}{2} (\a_2\del_0-\a_3\del_1-\a_0\del_2+\a_1\del_3)\cr
&\hat p_3=\frac{i}{2} (\a_3\del_0+\a_2\del_1-\a_1\del_2-\a_0\del_3).}
\eqno(2.7)
$$

\ni Defining the operator $\hat V$ as multiplication by the
matrix $V$ therefore leads to the commutation relations

$$
\eqalign{
&[\hat V_A{}^B,\hat V_C{}^D]=0\cr
&[\hat  p_i,\hat V_A{}^C]=-\frac{i}{2}\, \tau_{iA}{}^B\hat V_B{}^C\cr
&[\hat p_i,\hat p_j]=i\,\e_{ijk}\, \hat p_k,}\eqno(2.8)
$$

\ni which are the quantum equivalents of the
Poisson brackets of the corresponding classical quantities).
These commutation relations are familiar from the
Hamiltonian $SU(2)$-lattice gauge theory [11], where the $\hat p_i$ are
hermitian operators. Here, in contrast, we shall associate three {\it
complex} (i.e. $SL(2,\C)$) degrees of freedom with each link of the
hypercubic lattice. As usual, operators $(\hat V,\hat p)$ associated
with different links commute.

The entire construction (2.1-8) makes sense
also if we complexify the  group to $SU(2)_\C=SL(2,\C)$. We will
denote the corresponding group parameters by complex numbers
$\a_i^\C$, $i=0\dots 3$, again subject to a condition $\sum_{i=0}^3
(\a_i^\C)^2=1$.  The operators $\hat V$ and $\hat p$ are taken to
act on a space of holomorphic wave functions on the group manifold
of $SL(2,\C)$, whose inner product will be specified below. Note that
due to the non-compactness and non-abelianness of the group
$SL(2,\C)$, there is no  analogue of the bi-invariant Haar measure
$dg$, which exists on $SU(2)$.
However, thanks to  the work of Hall [12], we know
there exist unitary holomorphic transforms from  the space
$L^2(SU(2),dg)$ of square-integrable  functions on $SU(2)$ to spaces
$L^2(SL(2,\C),d\nu)^{\cal H}$ of holomorphic and
$\nu$-square-integrable functions on $SL(2,\C)$. (These are analogous
to the Segal-Bargmann integral transform from $L^2(\R^n)$ into the
holomorphic functions on $\C^n$.) This provides us with the desired
scalar product on functions on a complex group manifold. Note that
Hall's results have recently been used to construct a coherent state
transform for spaces of connections in the continuum [13].

We now recall some details of Hall's construction insofar as they are
relevant to the present discussion. For each real $t>0$, there is a
``coherent-state transform" $C_t:L^2(SU(2),dg)\rightarrow
L^2(SL(2,\C),d\nu_t)^{\cal H}$ defined by

$$
[C_t(f)](g_\C):=\int_{SU(2)} f(g)\rho_t(g^{-1}g_\C)\, dg,\eqno(2.9)
$$

\ni where $f\in L^2(SU(2),dg)$, $g\in SU(2)$, $g_\C\in SL(2,\C)$,
and $\rho_t$ is the heat kernel for the Casimir operator
$\Delta=-4\sum_i \hat p_i^2$ on $SU(2)$, i.e. the fundamental solution
at the identity of the heat equation $d\rho/dt =\frac12\Delta\rho$.
More precisely, since the argument of $\rho_t$ in (2.9) is a complex
group element, we are using its analytic continuation, which is
well-defined (see [12] for details). In terms of the explicit
parametrization (2.1) for the matrices $V(g)$ and the normalized Haar
measure $dg$, and using a series expansion for the heat kernel, one
obtains

$$
\eqalign{
[C_t(f)](g_\C)=
&\frac{1}{\pi^2}\int d\a_0\int
d\a_1\int d\a_2 \int d\a_3 \, \d (\sum_{i=0}^3 a_i^2-1)\,
f(\a_i)\,\times \cr
&\sum_{j=0,\frac12,\dots} (2 j+1) \, e^{-j (j+1) t/2}\,
U_{2 j}(a_0 a^\C_0 +a_1 a^\C_1 +a_2 a^\C_2 +a_3 a^\C_3)=\cr
\frac{1}{\pi^2}\int\limits_{-1}^1 d\a_0&\int\limits_{-\sqrt{1-a_0^2}}^
{\sqrt{1-a_0^2}}
d\a_1\int\limits_{-\sqrt{1-a_0^2-a_1^2}}^{\sqrt{1-a_0^2-a_1^2}} d\a_2
\frac{1}{2\sqrt{1-a_0^2-a_1^2-a_2^2}}
 \sum_{j=0,\frac12,\dots} (2 j+1)\, \times \cr
 &e^{-j (j+1) t/2} ((f(\a_i)
U_{2j})|_{a_3=\sqrt{1-a_0^2-a_1^2-a_2^2}}+ (f(\a_i)
U_{2j})|_{a_3=-\sqrt{1-a_0^2-a_1^2-a_2^2}})}\eqno(2.10)
$$

\ni (whenever the infinite sum over $j$ on the right-hand side
converges), where the
$U_{2j}$ denote the Chebyshev polynomials of the second kind. The
image of a square-integrable function $f(g)$ on $SU(2)$ is a
holomorphic function on $SL(2,\C)$, square-integrable with respect to
$d\nu_t$, which is essentially the heat kernel measure on the quotient
space $SL(2,\C)/SU(2)$.

Since our aim is a manifestly gauge-invariant description of lattice
gravity in a holomorphic loop representation, several issues have to be
resolved. Firstly, we are not aware of a simple explicit expression
for the measure $d\nu_t$ in the holomorphic representation obtained
via (2.9), and therefore have to look for functions $f(g)$ with a
simple transformation behaviour, preferably elements of an orthonormal
basis of $L^2(SU(2),dg)$, which will be mapped into orthonormal
functions of $L^2(SL(2,\C),d\nu_t)^{\cal H}$, since $C_t$ preserves
scalar products. Secondly, we will work with functions
that are gauge scalars and can be expressed as functions
of Wilson loops, i.e. traced holonomies of closed
loops on the lattice. The lattice
Hamiltonian maps such functions into themselves.

To obtain the holomorphic transform for general $SU(2)$-wave
functions on the lattice, we need to take the product over all
lattice links of the transform for a single link, formula (2.10).
First however we will give an (overcomplete) set of functions of a
single copy of $SU(2)$, i.e. on a single link, that have a simple
transformation behaviour under the transform (2.10). They are given by
appropriate sums of polynomials in the four real parameters $\a_i$
(restricted to the submanifold $SU(2)\subset \R^4$) and can be labelled
by the exponents $n_i$ in the polynomial $\a_0^{n_0} \a_1^{n_1}
\a_2^{n_2} \a_3^{n_3}$ of highest order occurring in the sum. One finds

$$
\eqalign{
&p(n_0,n_1,n_2,n_3):=\cr
&\sum_{j_0=1}^{[n_0/2+1]}
\frac{ (-1)^{j_0-1}\, a_0^{ n_0-2 ( j_0-1)} }
{ 2^{2 (j_0-1)}\, (j_0-1)!\, (n_0-2 (j_0-1))!}
\frac{ (n_0)! (n_0-j_0+n_1+n_2+n_3+1)!}
{(n_0+n_1+n_2+n_3)!}\cr
&\sum_{j_1=1}^{[n_1/2+1]}
\frac{ (-1)^{j_1-1}\, a_1^{ n_1-2 (j_1-1)} }
{ 2^{2 (j_1-1)}\, (j_1-1)!(n_1-2 (j_1-1))!}
\frac{  (n_1)! (n_0-j_0+n_1-j_1+n_2+n_3+2)!}
{(n_0-j_0+n_1+n_2+n_3+1)!}\cr
&\sum_{j_2=1}^{[n_2/2+1]}
\frac{ (-1)^{j_2-1}\, a_2^{ n_2-2 (j_2-1)} }
{ 2^{2 (j_2-1)}\, (j_2-1)!(n_2-2 (j_2-1))!}
\frac{  (n2)! (n_0-j_0+n_1-j_1+n_2-j_2+n_3+3)!}
{(n_0-j_0+n_1-j_1+n_2+n_3+2)!}\cr
&\sum_{j_3=1}^{[n_3/2+1]}
\frac{ (-1)^{j_3-1}\, a_3^{ n_3-2 (j_3-1)} }
{ 2^{2 (j_3-1)}\, (j_3-1)! (n_3-2 (j_3-1))!}
\frac{ (n_3)!(n_0-j_0+n_1-j_1+n_2-j_2+n_3-j_3+4)!}
{(n_0-j_0+n_1-j_1+n_2-j_2+n_3+3)!}.}\eqno(2.11)
$$

\ni For the holomorphic transform of $p(n_0,n_1,n_2,n_3)$ one obtains

$$
[C_t(p(n_0,n_1,n_2,n_3))](\a_i^\C)=e^{-(n_0+n_1+n_2+n_3)
(n_0+n_1+n_2+n_3+2)t/8} p(n_0,n_1,n_2,n_3)^\C,\eqno(2.12)
$$

\ni where by $p(n_0,n_1,n_2,n_3)^\C$ we denote the expression (2.11)
with the real parameters $\a_i$ replaced by the corresponding complex
quantities $\a_i^\C$. That is, to find the (inverse) holomorphic
transform of a polynomial function of the $\a_i$ ($\a_i^\C$), one first
has to express it as a linear combination of the $p(n_0,n_1,n_2,n_3)$
($p(n_0,n_1,n_2,n_3)^\C$) and then use (2.12). Both (2.11) and (2.12)
are crucial formulas for relating the $SU(2)$- and the holomorphic
$SL(2,C)$-representation, and will be used in the
following sections. The next step is the identification of
gauge-invariant combinations of the $p(n_0,n_1,n_2,n_3)$. If the
configuration space consisted of just one link (with endpoints
identified), our task would be straightforward: all gauge-invariant
quantities one can construct in that case are functions of $\frac12
{\rm Tr}\,V= \a_0$, and a complete basis is given by the Chebyshev
polynomials $U_{n_0}(\a_0)$.

The situation on the hypercubic lattice is
more complicated since general gauge-invariant quantities are
functions of traces of holonomies around arbitrary lattice loops,
which do not necessarily factorize into products of link contributions.
Moreover, there is the additional problem of finding a set of
gauge-invariant functions on $\times_l SU(2)$ that is complete (i.e.
spans the Hilbert space of square-integrable functions), but at the
same time not overcomplete, i.e. free of redundant degrees of freedom.
This is a well-known complication with intrinsically gauge-invariant
formulations of lattice gauge theory, and there are various strategies
of dealing with it. In the section 4 we will address some of these
difficulties in the context of the $1\times 1\times 1$-lattice.

Formula (2.11) can be used to construct gauge-invariant
functions with a simple transformation behaviour that are labelled by
lattice loops. This is important because the (overcomplete) basis
of gauge-invariant functions $\{ {\rm Tr}\,V_{i_1}V_{i_2}\dots
V_{i_n},\quad \g=l_{i_1}\circ l_{i_2}\circ\dots\circ l_{i_n}\;{\rm
a\;lattice\; loop}\}$ often appears in applications. Using the
parametrization (2.1) for the link matrices $V_{i}$, ${\rm
Tr}\,V_{i_1}V_{i_2}\dots V_{i_n}$ is a homogeneous sum of polynomials
in those parameters. Re-expressing in each summand the link
contributions in terms of the functions $p(n_0,n_1,n_2,n_3)$, one
obtains quantities that transforms like (2.12) with $t$-dependent
exponential factors.

\vskip2cm

\line{\ch 3 Solutions to the Wheeler-DeWitt equation\hfil}

Next we study the action of a discretized form of the phase space
Hamiltonian for canonical gravity. Let us label lattice sites by
an integer $n$ and the three positive directions emanating from each
site by $\hat a=\hat 1,\hat 2$ or $\hat 3$. Thus the canonical
variables are given by $V(n,\hat a)$ and $p(n,\hat a)$. We denote the
holonomy of a plaquette loop based at the site $n$ in the
$\hat a$-$\hat b$-plane by $V(n,P_{\hat a\hat b})$, that is,
$V(n,P_{\hat a\hat b})= V(n,\hat a)V(n+\hat a,\hat b)V(n+\hat b,\hat
a)^{-1}V(n,\hat b)^{-1}$.

We require that in the limit as all link lengths go to
zero the continuum Hamiltonian is recovered. Since our lattice was
assumed to be purely topological, we define this limit with respect
to an auxiliary Euclidean coordinate system (the three lattice
directions coinciding with the three coordinate axes) in which all
links have length $a$. As $a\rightarrow 0$, one derives the usual
expansion for the plaquette holonomy

$$
 V(n,P_{\hat a\hat b})_A{}^B
\buildrel{a\rightarrow 0}\over\longrightarrow  \one_A{}^B + a^2
F_{ab}^k \tau_{kA}{}^B + O(a^3),\eqno(3.1)
$$

\ni where $F_{ab}$ is the $a$-$b$-component of the field strength
associated with the selfdual connection $A$. For the momentum
variable $p_i(n,\hat a)$ we require that

$$
p_i(n,\hat a) \buildrel{a\rightarrow 0}\over\longrightarrow  a^2
\tilde E^a_i(n) +O(a^3)\eqno(3.2)
$$

\ni in terms of the continuum momentum density $\tilde E$, since $p$
is like a momentum variable smeared in one (out of three) spatial
directions. Therefore, if we associate with each lattice site $n$
the lattice Hamiltonian

$$
H^\C(n)=\sum_{\hat a <\hat b}\e^{ijk} p_i(n,\hat a) p_j(n,\hat b)
{\rm Tr}\, (V(n,P_{\hat a \hat b})\tau_k),\eqno(3.3)
$$

\ni this to lowest order leads, up to a power of $a$, to the correct
continuum limit

$$
H^\C(n) \buildrel{a\rightarrow 0}\over\longrightarrow  a^6 \e^{ijk}
\tilde E_i^a \tilde E_j^b F_{ab\, k} + O(a^7).\eqno(3.4)
$$

\ni The total Hamiltonian is given by the sum $\sum_n H^\C(n)$.
As a result of the ``dimensional transmutation", the
discretized Hamiltonian is a topological quantity, in the sense
that it depends only on link variables, with no reference to the
link length $a$, and acts on loop wave functions in a combinatorial
way. As long as we do not re-introduce a length scale, the formulation
is therefore purely topological.

Just as in lattice gauge theory, the requirement of the correct
continuum limit does not fix the discretized Hamiltonian uniquely. In
the present work we do not investigate the question whether a
different choice of $H^\C(n)$ leads to equivalent results. In going to
the quantum theory, another ambiguity arises in the choice of the
operator ordering of $\hat H^\C(n)$. The most commonly used operator
ordering for the Hamiltonian is the one
with the operators $\hat p_i$ to the right,
but also the opposite ordering with both of the $\hat p_i$ to the left
is sometimes used [14]. In our present investigation we will be using
the former, i.e.

$$
\hat H^\C =\sum_n \sum_{\hat a <\hat b}
e^{ijk} {\rm Tr}\, (\hat V(n,P_{\hat a
\hat b})\tau_k) \, \hat p_i(n,\hat a)\, \hat p_j(n,\hat b).
\eqno(3.5)
$$

Since the spatial diffeomorphisms have been taken care of, the only
remaining task is to look for holomorphic wave functions
$\psi^\C\in\times_l L^2(SL(2,\C),d\nu_t)^{\cal H}$ that solve the
analogue of the Wheeler-DeWitt equation,

$$
\hat H^\C \psi^\C =0.\eqno(3.6)
$$

The existence of solutions depends on the spectral properties of the
Hamiltonian operator $\hat H^\C$. Since the constituent operators
$\hat V$ and $\hat p$ are not selfadjoint in the holomorphic
representation, we do not have any a-priori information about
the spectrum of $\hat H^\C$. There are three possibilities:
\item{(i)} $\hat H^\C\psi^\C=0$ has non-trivial solutions in
$\times_l L^2(SL(2,\C),d\nu_t)^{\cal H}$;
\item{(ii)} $\hat H^\C\psi^\C=0$ does not have solutions in
$\times_l L^2(SL(2,\C),d\nu_t)^{\cal H}$, but can be solved ``as
a differential equation", i.e. there are non-square-integrable
solutions;
\item{(iii)} there are no solutions.

\ni In case (i), the solution space inherits a scalar
product from the original Hilbert space, whereas in case (ii)
one still has to define a suitable inner product. - A number of
identities are useful in computing the action of the Hamiltonian on
gauge-invariant wave functions (which all contain terms of the form
${\rm Tr}\, V_{i_1}V_{i_2}\dots$). The first one is

$$
\e^{ijk}\tau_{j\,A}{}^B\tau_{k\,C}{}^D=
\tau_{i\,A}{}^D\d_C{}^B- \tau_{i\,C}{}^B\d_A{}^D,\eqno(3.7)
$$

\ni from which follow two identities for the traces of holonomies,

$$
\eqalign{
&\e^{ijk} {\rm Tr}\, (V_\a\tau_j V_\b\tau_k)=
{\rm Tr}\,V_\b {\rm Tr}\, (V_\a\tau_i)-
{\rm Tr}\,V_\a {\rm Tr}\,(V_\b \tau_i)\cr
&\e^{ijk} {\rm Tr}\, (V_\a\tau_j) {\rm Tr}\,(V_\b\tau_k)=
{\rm Tr}\, (V_\a\tau_i V_\b )-
{\rm Tr}\, (V_\a V_\b\tau_i),}\eqno(3.8)
$$

\ni where $\a$, $\b$ are two lattice loops intersecting at $n$, and
the product loop $\a\circ\b$ is obtained by the usual loop composition
at $n$. Secondly, we have

$$
\tau_{i\, A}{}^B\tau_{i\,C}{}^D =\d_A{}^B\d_C{}^D-
2\,\d_A{}^D\d_B{}^C.\eqno(3.9)
$$

\ni Lastly, there is the well-known identity relating a product of
two Wilson loops of $SL(2,\C)$-holonomies to a sum of two Wilson
loops,

$$
{\rm Tr}\,V_\a {\rm Tr}\,V_\b ={\rm Tr}\,V_{\a\circ\b}+{\rm
Tr}\,V_{\a\circ\b^{-1}}\equiv {\rm Tr}\,V_\a V_\b +{\rm Tr}\,V_\a
V_\b^{-1}.\eqno(3.10)
$$

\ni By virtue of these identities, the action of the
Hamiltonian on states of the form ${\rm Tr}\, V_{i_1}V_{i_2}\dots$ may
be interpreted as cutting and joining of the lattice loop arguments,
as is typical for the loop representation. To what extent such a
geometric interpretation is useful in finding solutions to the
zero-eigenvalue equation depends to some extent on the type of basis
chosen for the quantum states. As in the case of Hamiltonian lattice
gauge theory, the Hamiltonian couples neighbouring links due to the
occurrence of the plaquette holonomy operators $\hat V(n,P_{\hat a\hat
b})$ in $\hat H(n)$.

Next we show that solutions to the zero-eigenvalue equation
$\hat H^\C\psi^\C=0$ are in one-to-one correspondence with solutions
to $\hat H_{ind}\psi^\C=0$, where $\hat H_{ind}$ is the
(self-adjoint) Hamiltonian induced from the $SU(2)$-lattice theory.
That is, we take $\hat H^\C$ as in (3.5), and substitute the
operators by their real counterparts acting on $\times_l
L^2(SU(2),dg)$ (i.e. take all parameters $\a_i$ etc. to be real),
before translating it to the holomorphic
representation using the transform (2.10).

Let us for the moment assume we are given a complete orthogonal basis
of wave functions
$\{\chi (\vec n),\quad \vec n=(n_1,n_2,\dots n_d)\}$, $d={\rm
dim} (\times_l SU(2)/\times_s SU(2) )$, for the gauge-invariant(real)
Hilbert space $L^2(\times_l SU(2)/\times_s SU(2), \pi (\prod_l dg))$,
where $ \pi (\prod_l dg)$ denotes the projection to the quotient
space of the product of Haar measures, $l$ is the number of links and
$s$ the number of sites. Its elements are labelled by
integers $n_i$ and transform according to

$$
C_t(\chi (\vec n))=e^{-f_t(\vec n)} \chi^\C (\vec n),\eqno(3.11)
$$

\ni and $f_t(\vec n)\geq 0$. Since the exponential factors are just
real numbers rescaling the basis (which we did not assume to be
ortho{\it normal}), we can write any square-integrable element of the
holomorphic Hilbert space as a real linear combination of the $\chi^\C
(\vec n)$, and have $<\chi^\C (\vec m),\chi^\C (\vec n)>\sim
\d_{\vec m,\vec n}$. The action of the Hamiltonian $\hat H^\C$ acting
on a general vector $\sum_{\vec n} a(\vec n)\,\chi^\C (\vec n)$,
$a(\vec n)\in\R$, can be written as a matrix equation

$$
\hat H^\C \sum_{\vec n} a(\vec n)\,\chi^\C (\vec n)=
\sum_{\vec n}\sum_{\vec m} M(\vec n,\vec m)\, a(\vec n)\,\chi^\C (\vec
m).\eqno(3.12)
$$

\ni The Wheeler-DeWitt equation (3.6) is therefore equivalent to an
infinite tower of equations for the coefficients $a(\vec n)$,

$$
\sum_{\vec n} M(\vec n,\vec m)\, a(\vec n)=0,\quad \forall\vec m.
\eqno(3.13)
$$

The norm of a vector $\sum_{\vec n} a(\vec n)\,\chi^\C (\vec n)$ can
be calculated using its inverse image under (3.11), and is given by

$$
||\sum_{\vec n} a(\vec n)\,\chi^\C (\vec n)||_{SL(2,\C)} =
\sqrt{\sum_{\vec n} a(\vec n)^2\, e^{2 f_t(\vec n)}\, ||\chi(\vec
n)||^2_{SU(2)} }.\eqno(3.14)
$$

By contrast, the  Hamiltonian $\hat
H^\C_{ind}$ acts on holomorphic wave functions according to

$$
\hat H^\C_{ind} \sum_{\vec n} b(\vec n)\,\chi^\C (\vec n)=
\sum_{\vec n}\sum_{\vec m} e^{-f_t(\vec m)} \,M(\vec n,\vec m)\,
e^{f_t(\vec n)}\, b(\vec n)\,\chi^\C (\vec m),\eqno(3.15)
$$

\ni leading to the set of conditions

$$
\sum_{\vec n} M(\vec n,\vec m)\, e^{f_t(\vec n)}\, b(\vec n)=0,\quad
\forall\vec m\eqno(3.16)
$$

\ni on the coefficients $b(\vec n)$ of zero-eigenvectors. Clearly
they are related to the solutions of (3.13) by $a(\vec n)=e^{f_t(\vec
n)} b(\vec n)$. However, the norm of the solution vector is in general
different, and we have

$$
||\sum_{\vec n} b(\vec n)\,\chi^\C (\vec n)||_{SL(2,\C)} =
\sqrt{\sum_{\vec n} b(\vec n)^2\, e^{2 f_t(\vec n)}\, ||\chi(\vec
n)||^2_{SU(2)}}\leq ||\sum_{\vec n} a(\vec n)\,\chi^\C (\vec
n)\,||_{SL(2,\C)} .\eqno(3.17)
$$

\ni It may therefore happen that a zero-eigenvalue solution to $\hat
H^\C_{ind}\sum_{\vec n} a(\vec n)\,\chi^\C (\vec n)=0$ is
square-integrable, while the corresponding solution to
$\hat H^\C \sum_{\vec n} a(\vec n)\,\chi^\C (\vec n)=0$ is not.

The form of the quantum Hamiltonian (3.5), like its counterpart in the
continuum theory, is sufficiently complicated so as not to lead
us to expect the eigenvalue problem could be soluble trivially. What
comes to help in the analogous problem in the continuum is the
existence of ``algebraically special" solutions. For example, in the
representation where  $H_{cont}\sim \e^{ijk}\tilde E_i^a \tilde E_j^b
F_{ab\, k}$ is quantized by $\hat H_{cont}\sim \e^{ijk} F_{ab\,
k}(A)\,\del^2/ \del A^i_a \del A^j_b$, solutions are given by wave
functions $\psi(A,\g):={\rm Tr}\, P\exp\oint_\g
A_a(\g(t))\dot\g^a(t)\, dt$, whenever $\g$ is a smooth,
non-selfintersecting loop [15]. This happens because the derivatives
$\del/\del A$ bring down two factors of the tangent vector,
$\dot\g^a\dot\g^b$, which vanish when multiplied by the antisymmetric
tensor $F_{ab}$.

It turns out that there exist analogous solutions in the lattice
formulation. Take any straight ``Polyakov loop" $\a$, i.e. a loop
without corners that winds around the lattice once, and is therefore
non-contractible. On a $N^3$-lattice this is a loop $\a$ made up of $N$
consecutive links in a given direction $\hat a$, $\hat b$ or $\hat c$
(Fig.1). The contribution of $\hat H^\C (n)$ to the Hamiltonian acting
on a wave function $\sim {\rm Tr}\, V_\a$ vanishes at any given site
$n$ crossed by $\a$, since it would need a wave function with support
in at least two independent lattice directions to be non-zero.

\epsffile{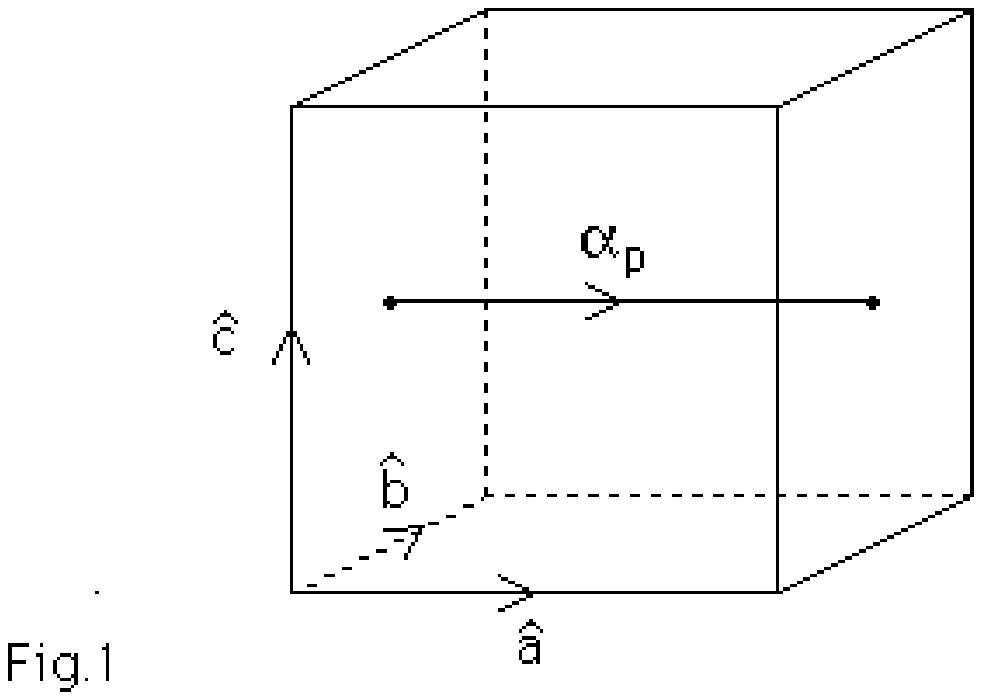}

There are $3N^2$ such Polyakov loops (which we will take to be
positively oriented), $N^2$ in each direction. They will be denoted by
a subscript `p'. The Polyakov loops can be parametrized by three
integers, namely the coordinates of the point where they intersect
one of the three planes $\hat a=0$, $\hat b=0$ or $\hat c=0$. (Note
that each lattice link is  contained in exactly one Polyakov loop.)
We denote the corresponding wave functions by $\phi(\a_p)\equiv {\rm
Tr}\,V_{\a_p}$. It is easy to see that any wave function
$\phi(\a_p^n)$ of a multiple Polyakov loop
$\a_p^n=\a\circ\a\dots\circ\a$ with winding number $n$ is also
annihilated by $\hat H^\C$. The solution space to the Wheeler-DeWitt
equation is therefore infinite-dimensional. However, because of the
non-linearity of the Hamiltonian, it is in general not true that $\hat
H^\C\phi(\a_p^n)\phi(\b_p^m)$ vanishes too. This only occurs when
$(\a_p,\b_p)$ are a pair of non-intersecting Polyakov loops. We have
therefore found: any wave function that is a linear combination of
terms in

$$
\{ \phi(\a_p^{n_1})
\phi(\b_p^{n_2})\dots
\phi(\omega_p^{n_k}),\quad
\a_p,\b_p,\dots\omega_p\;\,k\;{\rm non-intersecting \;Polyakov\;
loops\} }\eqno(3.18)
$$

\ni is annihilated by the discretized Wheeler-DeWitt operator $\hat
H^\C$. The space of such functions is a rather large linear subspace
of the original holomorphic Hilbert space. Characteristically, its
elements are all highly non-local wave functions on the lattice. Note
that we need not consider separately products of the form $\phi(\a_p^m)
\phi(\a_p^n)$ for multiples of the same Polyakov loop $\a_p$,
since these can always be re-expressed via the trace identity (3.10) as
sums of elements of (3.18). Whether there exist
solutions that are not of this form remains to be explored (see
also the discussion in the next section).

{}From the point of view of non-perturbative quantum gravity, one is
interested in the structure of the solution space to $H^\C\psi^\C=0$,
in particular its scalar product and natural self-adjoint operators
acting on it. For the subsector of Polyakov wave functions described
above, there is an induced scalar product from the original Hilbert
space of holomorphic wave functions. In deriving this inner
product, an ambiguity arises because the scalar product on the wave
functions obtained through the holomorphic transform (2.10) is by
construction invariant under
right and left multiplication by $SU(2)$-matrices, but {\it not}
bi-invariant under $SL(2,\C)$. However, on the explicitly
($SL(2,\C)$-)gauge-invariant space of Polyakov wave functions
there is only a small remnant of this gauge covariance:
it turns out that the norm of a complex wave function depends on the
number of link variables that appear in the coordinate expression for
$\phi(\a_p^n)$.

This is to be contrasted with the pure $SU(2)$-case, say. There, in
order to simplify the computation of scalar products, one often uses a
gauge-fixing for a maximal number of link variables, which therefore
do not any more appear in the calculation. The right- and
left-invariance of the Haar measure ensures that no physical
quantities are affected by this choice (see, for example, [16]). In the
present case of $SL(2,\C)$, one may also introduce a gauge-fixing for
some of the links, but one has to keep track of it. Different
gauge-fixings result in a rescaling of the wave functions. The norm of
a Polyakov wave function $\phi(\a_p)$ on a $N^3$-lattice  without any
gauge-fixing can be computed using (2.11,12), and is found to
be

$$
||\phi(\a_p)||_{SL(2,\C)}=e^{3Nt/8}.\eqno(3.19)
$$

\ni If $m$ of the $N$ links occurring in $\a_p$ are
gauge-fixed, the norm changes accordingly to $e^{3(N-m)t/8}$.
Scalar products between (multiple) Polyakov wave functions
$\phi(\a_p^m)$ and $\phi(\b_p^n)$
vanish whenever the underlying Polyakov loops $\a_p$ and $\b_p$ are
different.  Furthermore one finds that for fixed Polyakov loop
$\a_p$, different multiples $\phi(\a_p^m)$ and $\phi(\a_p^n)$ generally
have a non-zero scalar product, but
do not form an orthogonal set, i.e. $<\phi(\a_p^m),\phi(\a_p^n)>\not=
\d_{mn}$.
Their norms and scalar products can easily be computed using (2.11) and
(2.12).

The next step in the investigation is the search for selfadjoint
operators, acting on the Hilbert space of the Polyakov wave functions.
Natural candidates are the holomorphic transforms of selfadjoint
operators in the $SU(2)$-representation that map Polyakov wave
functions into themselves. Of course one has to make sure that
the final physical expectation values do not
depend on a particular gauge-fixing or on the auxiliary
parameter $t$. Our lattice formulation does for the first time permit
rigorous questions about the physical observables and their operator
spectra. A detailed study of these issues will appear elsewhere.

\vskip2cm

\line{\ch 4 Gravity on the $1\times 1\times 1$-lattice\hfil}

To illustrate the technicalities involved in the search for solutions
other than the Polyakov wave functions of the previous section,
we now turn to the case of the $1\times 1\times 1$-lattice with
periodic boundary conditions. There is a single site $s$ and three
(oriented) links emanating from it, which we call $\a$, $\b$ and
$\g$ (Fig.2), to be identified with the $\hat 1$-, $\hat 2$- and
$\hat 3$-directions on the lattice.

\epsffile{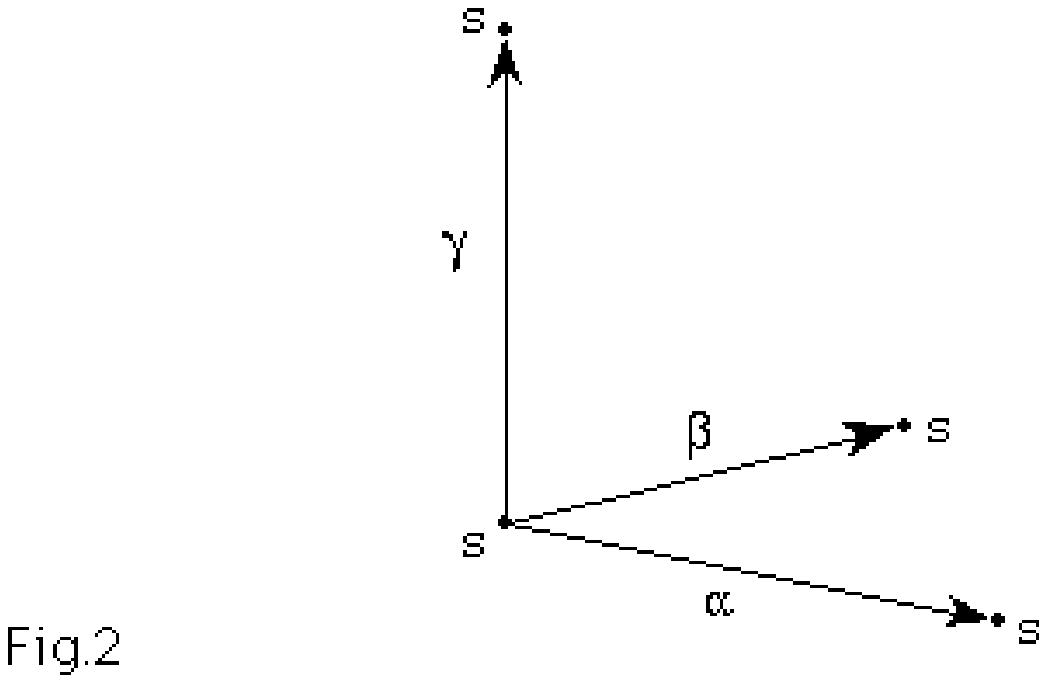}

We first discuss the gauge-invariant Hilbert space for the
$SU(2)$-theory and then apply the holomorphic transform (2.9,10).
By virtue of the boundary conditions, $\a$, $\b$ and $\g$ are
themselves closed loops. The corresponding holonomy matrices are
parametrized according to (2.1) by real parameters $\a_i$, $\b_i$ and
$\g_i$ respectively. The gauge transformations take values in a single
copy of the gauge group $SU(2)$, located at the site $s$. For this
case we have complete control over the gauge-invariant functions, i.e.
we can give a complete, non-redundant basis of the physical Hilbert
space. Following [18], a good set of local coordinates on the
six-dimensional physical configuration space $(SU(2)_\a\times SU(2)_\b
\times SU(2)_\g)/SU(2)_s$ is given by

$$
\eqalign{
&L_1(\a)=\frac12 {\rm Tr}\,V_\a =\a_0 \cr
&L_1(\b)=\frac12 {\rm Tr}\,V_\b =\b_0 \cr
&L_1(\g)=\frac12 {\rm Tr}\,V_\g =\g_0 \cr
&L_2(\a,\b)=\frac14 ({\rm Tr}\,V_\a V_\b^{-1}-{\rm Tr}\,V_\a V_\b)=
\vec\a\cdot\vec\b\cr
&L_2(\a,\g)=\frac14 ({\rm Tr}\,V_\a V_\g^{-1}-{\rm Tr}\,V_\a V_\g)=
\vec\a\cdot\vec\g\cr
&L_2(\b,\g)=\frac14 ({\rm Tr}\,V_\b V_\g^{-1}-{\rm Tr}\,V_\b V_\g)=
\vec\b\cdot\vec\g,}\eqno(4.1)
$$

\ni where we have used a vectorial notation for the ``spatial"
parameters of the holonomy matrices, $\vec\a :=(\a_1,\a_2,\a_3)$ etc.
In order to obtain a good global parametrization, one needs to add a
discrete parameter, which we take to be

$$
{\rm sign}(L_3(\a,\b,\g))={\rm sign}(\vec\a\times\vec\b\cdot\vec\g).
\eqno(4.2)
$$

\ni Also $L_3$ may be written as a linear combination of traces of
holonomies involving $V_\a$, $V_\b$ and $V_\g$ [17]. The need for such
discrete variables has been repeatedly emphasized by Watson [18].
Roughly speaking, the gauge-invariant Hilbert space of
square-integrable functions, $L^2(SU(2)_\a\times SU(2)_\b\times
SU(2)_\g)/SU(2)_s,\pi (\prod_l dg))$ of the quantum theory is spanned
by polynomials in the six quantities (4.1) and the discrete variable
by a copy of $\Z_2$.  For making contact with the corresponding
$SL(2,\C)$-representation, we have to form a Chebyshev basis from
these polynomials, which is orthogonal and has a simple transformation
behaviour under (2.10).

For simplicity, we will first describe the subtheory
obtained by reduction to the $\hat 1$-$\hat 2$-plane, say. The
relevant classical variables are then $L_1(\a)$, $L_1(\b)$ and
$L_2(\a,\b)$. Since the real polynomials $\a_0^{n_1} \b_0^{n_2}
(\vec\a\cdot\vec\b)^{n_3}$, regarded as elements of the physical
Hilbert space, do not have a straightforward holomorphic transform, we
use formula (2.11) to identify appropriate linear
combinations of such polynomials, in order to establish the explicit
relation between the $SU(2)$- and the holomorphic
$SL(2,\C)$-representation. For those, one finds

$$
\eqalign{
&eig(n_1,n_2,n_3):= \sum_{j_0=1}^{[\frac{n_1}{2}+1]}
\frac{(-1)^{j_0-1}}{2^{2 (j_0-1) }}
\frac{ n_1! (n_1-j_0+n_3+1)!}{ (j_0-1)! (n_1+n_3)! (n_1 -2 j_0+2)!}\cr
&\sum_{k_0=1}^{[\frac{n_2}{2}+1]}
\frac{(-1)^{k_0-1}}{2^{2 (k_0-1) }}
\frac{ n_2! (n_2-k_0+n_3+1)!}{ (k_0-1)! (n_2+n_3)! (n_2 -2 k_0+2)!}\;
\sum_{i=0}^{[\frac{n_3}{2}]} \sum_{k=0}^{i}
\sum_{l=0}^{i}\cr
&\left( \sum_{k_1=0}^{k}\sum_{l_1=0}^{l}
(-1)^{k_1} \left(\matrix{k\cr k1}\right)
(-1)^{l_1} \left(\matrix{l\cr l1}\right)
a_0^{n_1-2 j_0+2+2 k_1} b_0^{n_2-2 k_0+2+2 l_1}
(\vec\a\cdot\vec\b)^{n_3-2 i}\right) \cr
&\prod_{p=0}^{2 i-1} (n_3-p)\;
 \prod_{q=0}^{i-l-1} \frac{1}{2 (n_2-k_0+n_3+1-q)}\;
 \prod_{r=0}^{i-k-1} \frac{1}{2 (n_1-j_0+n_3+1-r)}\cr
&\left(\matrix{i-k\cr i-k-l}\right)
\prod_{n=0}^{i-k-l-1} (2(n_3-i-n+\frac12 )(k+n+1))\;
\frac{(-1)^{k+1}}{2^i\, (i-l)! (i-k)!}, }\eqno(4.3)
$$

\ni where $n_i\geq 0$. To give a few simple examples, one has

$$
\eqalign{
&eig(0,0,0)=1\cr
&eig(1,0,0)=a_0\cr
&eig(2,0,0)=a_0^2-\frac14\cr
&eig(0,0,1)=\vec a\cdot\vec b\cr
&eig(0,0,2)=(\vec a\cdot\vec b)^2 +\frac{a_0^2}{4}+\frac{b_0^2}{4}-
\frac{5}{16}\cr
&eig(1,1,1)=a_0 b_0\, \vec a\cdot\vec b\cr
&eig(2,1,1)=a_0^2 b_0\, \vec a\cdot\vec b -\vec a\cdot\vec b\,
\frac{b_0}{6}\cr
&\dots\;.}\eqno(4.4)
$$

\ni Again, by construction, the highest-order
polynomial occurring in $eig(n_1,n_2,n_3)$ is $\a_0^{n_1} \b_0^{n_2}
(\vec\a\cdot\vec\b)^{n_3}$, and the remaining lower-order polynomials
in the sum ensure a simple transformation behaviour, which in this
case is given by

$$
C_t(eig(n_1,n_2,n_3)) = e^{-(n_1+n_3)(n_1+n_3+2)t/8}
e^{-(n_2+n_3)(n_2+n_3+2)t/8} eig(n_1,n_2,n_3)^\C.\eqno(4.5)
$$

It is computationally much simpler to determine the action of the
Hamiltonian $\hat H^\C$ on the non-orthogonal basis of holomorphic,
square-integrable wave functions $\{e(n_1,n_2,n_3)^\C\equiv
(\a_0^\C)^{n_1} (\b_0^\C)^{n_2} (\vec\a^\C\cdot \vec \b^\C)^{n_3}\}$,
look for zero-eigenvalue solutions and then use (4.3,5) for the
computation of scalar products, rather than apply $\hat H^\C$ on the
functions $eig(n_1,n_2,n_3)^\C$ directly. The quantum Hamiltonian for
the $1\times 1\times 1$-lattice consists of a single contribution $\hat
H^\C (n)$ and is

$$
\hat H^\C =\sum_{\hat a <\hat b}\e^{ijk} {\rm Tr}\, (\hat V(P_{\hat a
\hat b})\tau_k) \, \hat p_i(n,\hat a)\, \hat p_j(n,\hat b).\eqno(4.6)
$$

\ni The plaquette holonomies $ V(P_{\hat a \hat b})$ are defined using
the appropriate identifications of links, for example,
$ V(P_{\hat 1 \hat 2})=V_{\hat 1} V_{\hat 2} V_{\hat 1}^{-1} V_{\hat
2}^{-1}$.

We now describe our method for finding solutions to the equation
$\hat H^\C \sum_{\vec n} m(\vec n)\,e(\vec n)^\C=0$ for the
special case that the wave functions have support only
in the $\hat 1$-$\hat 2$-plane.
We first calculated the action of the Hamiltonian on a
basic holomorphic polynomial $e(n_1,n_2,n_3)^\C$, and obtained

$$
\eqalign{
\hat H^\C &e(n_1,n_2,n_3)^\C=
n_1 n_2\,(\, e(n_1-1,n_2-1,n_3+3)^\C- e(n_1-1,n_2-1,n_3+1)^\C\,)\cr
+&(n_1 n_2+n_1 n_3+n_2 n_3-n_3)\,(\, e(n_1,n_2,n_3)^\C-
 e(n_1,n_2,n_3+2)^\C\,) \cr
+&(n_1 n_2-n_1 n_3)\, e(n_1-1,n_2+1,n_3+1)^\C
+(n_1 n_2-n_2 n_3-)\,e(n_1+1,n_2-1,n_3+1)^\C\cr
+&(n_1 n_2+n_1 n_3+n_2 n_3-n_3^2)\,(\, e(n_1+2,n_2+2,n_3)^\C
-e(n_1+2,n_2,n_3)^\C\cr
-&e(n_1,n_2+2,n_3)^\C\,)\,
-(n_1 n_2-n_1 n_3-n_2 n_3-n_3^2-2 n_3)\,e(n_1+1,n_2+1,n_3+1)^\C\cr
+&n_1 n_3\,(\,e(n_1-1,n_2+1,n_3-1)^\C
-e(n_1-1,n_2+3,n_3-1)^\C\,)\cr
+&n_2 n_3\,(\,e(n_1+1,n_2-1,n_3-1)^\C
-e(n_1+3,n_2-1,n_3-1)^\C\,)\cr
-&(2 n_1 n_3+2 n_2 n_3+n_3^2+2 n_3)\,e(n_1+1,n_2+1,n_3-1)^\C\,
+(n_1 n_3+2 n_2 n_3 +n_3^2 +2 n_3)\cr
&e(n_1+3,n_2+1,n_3-1)^\C\,
+(2 n_1 n_3+n_2 n_3+n_3^2+2 n_3)\,e(n_1+1,n_2+3,n_3-1)^\C\cr
-&(n_1 n_3+n_2 n_3+n_3^2+2 n_3)\,e(n_1+3,n_2+3,n_3-1)^\C\cr
-&n_3\, (n_3-1)\,(\,e(n_1+2,n_2,n_3-2)^\C-e(n_1+4,n_2,n_3-2)^\C+
e(n_1,n_2+2,n_3-2)^\C\cr
-&3\, e(n_1+2,n_2+2,n_3-2)^\C+2\,e(n_1+4,n_2+2,n_3-2)^\C
-e(n_1,n_2+4,n_3-2)^\C\cr
+&2\, e(n_1+2,n_2+4,n_3-2)^\C-e(n_1+4,n_2+4,n_3-2)^\C\, ),
}\eqno(4.7)
$$

\ni for $n_i\geq 0$. That is, independent of the values of the
$n_i$, the right-hand side of (4.7) is a linear combination of (at
most) 26 terms, and the arguments of the $e(n_1,n_2,n_3)^\C$ occurring
lie in the ranges $-1\leq\Delta n_1\leq 4$, $-1\leq\Delta n_2\leq 4$
and $-2\leq\Delta n_3\leq 3$.

Since the functions $e(n_1,n_2,n_3)^\C$ form a (non-orthogonal) basis
for the gauge-invariant Hilbert space, we can reformulate the
zero-eigenvalue condition as an infinite set of equations for the
$m(n_1,n_2,n_3)$ obtained by setting the coefficient of each
$e(n_1,n_2,n_3)^\C$ on the right-hand side of $\sum_{n_i}\hat
H^\C m(n_1,n_2,n_3)\,e(n_1,n_2,n_3)^\C$ to zero. The general condition
can be labelled by three integers and is easily derived from (4.7),
yielding

$$
\eqalign{
C&[n_1,n_2,n_3]:=
(n_1+1)(n_2+1)\,(\,m(n_1+1,n_2+1,n_3-3)\,-m(n_1+1,n_2+1,n_3-1)\,)\cr
&-(n_2 n_3-n_1 n_2-n_1+n_3)\,m(n_1-1,n_2+1,n_3-1)\,
+(n_1 n_2+n_1 n_3+n_2 n_3-n_3)\cr
&m(n_1,n_2,n_3)\,
-((n_1-2) n_2+(n_1-2) n_3+n_2 n_3-n_3^2)\,m(n_1-2,n_2,n_3)\cr
&-(n_1 n_2+n_1 (n_3-2)+n_2 (n_3-2)-n_3+2)\,m(n_1,n_2,n_3-2)\cr
&-(-n_1 n_2+n_1 n_3-n_2+n_3)\,m(n_1+1,n_2-1,n_3-1)\cr
&-(n_1 (n_2-2)+n_1 n_3+(n_2-2) n_3-n_3^2)\,m(n_1,n_2-2,n_3)\cr
&-(n_1 n_2-n_1 n_3-n_2 n_3-n_3^2+2 n_3)\, m(n_1-1,n_2-1,n_3-1)\cr
&+(n_1+1)(n_3+1)\,(\,m(n_1+1,n_2-1,n_3+1)\,-m(n_1+1,n_2-3,n_3+1)\,)\cr
&+((n_1-2)(n_2-2)+(n_1-2)n_3+(n_2-2)n_3-n_3^2)\,m(n_1-2,n_2-2,n_3)\cr
&-(2 n_1 n_3+2 n_2 n_3+n_3^2+2 n_1+2 n_2-1)\,m(n_1-1,n_2-1,n_3+1)\,\cr
&+(n_1 n_3+2 n_2 n_3+n_3^2+n_1+2 n_2-n_3-2)\,m(n_1-3,n_2-1,n_3+1)\cr
&+(2 n_1 n_3+n_2 n_3+n_3^2+2 n_1+n_2-n_3-2)\,m(n_1-1,n_2-3,n_3+1)\cr
&-(n_1 n_3+n_2 n_3+n_3^2+n_1+n_2-2 n_3-3)\,m(n_1-3,n_2-3,n_3+1)\cr
&+(n_2+1)(n_3+1)\,(\,m(n_1-1,n_2+1,n_3+1)\,-m(n_1-3,n_2+1,n_3+1)\,)\cr
&-(n_3^2+3n_3+2)\,(\,m(n_1-2,n_2,n_3+2)\,-m(n_1-4,n_2,n_3+2)\,
 +m(n_1,n_2-2,n_3+2)\cr
&-3\,m(n_1-2,n_2-2,n_3+2)\,+2\,m(n_1-4,n_2-2,n_3+2)\,
 -m(n_1,n_2-4,n_3+2)\cr
&+2\,m(n_1-2,n_2-4,n_3+2)\,-m(n_1-4,n_2-4,n_3+2)\,)=0.}\eqno(4.8)
$$

Notice first that the two-dimensional wave functions
$e(n_1,n_2,n_3)^\C$ fall into two sectors which are mapped into
themselves under the action of the Hamiltonian. These are (i) the even
sector: either all $n_i$ are even or all $n_i$ are odd; (ii) the odd
sector: two $n_i$ are even and one $n_i$ is odd or vice versa. Hence
it suffices to investigate the two sectors separately.

The special solutions discussed in the previous section
correspond to all functions of the form $e(n_1,0,0)^\C$ and
$e(0,n_2,0)^\C$, and (4.8) imposes no conditions on the corresponding
coefficients $m(n_1,0,0)$ and $m(0,n_2,0)$. We will in the following
set these coefficients to zero, because we are interested in the
possible existence of other solutions. In order to tackle  the system
of equations in a well-defined manner, it is useful to define the
order $ord(C[n_1,n_2,n_3])$ by

$$
ord(C[n_1,n_2,n_3]):={\rm max}\{n_1+n_3,n_2+n_3\},\eqno(4.9)
$$

\ni and then try to solve simultaneously,
order by order, the sets of equations (4.8) of the same order,
eliminating coefficients $m(n_1,n_2,n_3)$ of higher order (with
`order' defined analogously to (4.9)) in terms of the lower order ones.
We have investigated the even sector and solved iteratively in this
manner the order-0, 2, 4, 6, 8, 10 and 12 equations.  There are 1
order-0 equation, 5 order-2, 13 order-4, 25 order-6, 41 order-8, 61
order-10 and 85 order-12 equations. Taking into account equations  up
to this order, one finds that for any solution to the Wheeler-DeWitt
equation the coefficients have to satisfy simultaneously

$$
\eqalign{
{\rm order\;2:}\,&m(0,0,2)=m(1,1,1)=m(2,2,0)=0\cr
{\rm order\;4:}\,&m(1,3,1)=m(3,1,1)=m(1,1,3)=m(2,0,2)=m(0,2,2)=
 m(0,0,4)=0\cr
&m(2,4,0)=m(4,2,0)=\frac{1}{9}\,m(2,2,2),\quad m(3,3,1)=\frac{38}{27}\,
m(2,2,2),\cr
&{\rm no\; conditions\; on\;}m(2,2,2),\; m(4,4,0)\cr
{\rm order\;6:}\,&m(1,5,1)=m(5,1,1)=-\frac{7}{9}\, m(2,2,2),\quad
 m(1,1,5)=\frac{2}{45}\, m(2,2,2),\cr
&m(4,0,2)=m(0,4,2)=-\frac{7}{18}\, m(2,2,2),\quad
 m(3,1,3)=m(1,3,3)=\frac{13}{27}\, m(2,2,2),\cr
&m(0,0,6)=-\frac{1}{135}\,m(2,2,2),\;\dots,}\eqno(4.10)
$$

\ni i.e. the coefficients $m$ of order 0 and 2 are completely fixed,
but at  order 4 there appear two free parameters, $m(2,2,2)$ and
$m(4,4,0)$. (Of course we could have solved in terms of other order-4
parameters.)  Unfortunately we cannot be sure whether the conditions
$C[n_1,n_2,n_3]=0$ for $ord(C)>12$ do not (through coupling among
equations of different order)
fix these parameters, although the behaviour
of the conditions evaluated so far does not make it seem likely. We
conjecture that there is an infinite number of free parameters (with
an increasing number of free parameters at each order), corresponding
to an infinite set of solutions to the discretized Wheeler-DeWitt
equation (beyond those coming from the Polyakov loops). These would
be of the form of (presumably infinite) linear combinations,
parametrized by those free parameters. Clearly then the question
arises of whether these solutions have a finite norm. Since we have
not even found a single explicit solution of this type, we are unable
to answer this question presently.

We will now have a brief look at the full $1\times 1\times 1$-lattice
theory. However, we will not attempt to solve the eigenvalue equation
directly, since the analogues of (4.7) and (4.8) contain about ten
times as many terms. The discussion is meant to serve as an
illustration of how to set up the gauge-invariant Hilbert space in a
symmetric way, and formulate the eigenvalue problem in principle. The
problem is the incorporation of the classical discrete degree of
freedom sign$(L_3(\a,\b,\g))$ in the quantum theory.  Although the
functions

$$
\{ \a_0^{n_1} \b_0^{n_2} \g_0^{n_3} (\vec\a\cdot\vec
\b)^{n_4} (\vec\a\cdot\vec\g)^{n_5} (\vec\b\cdot\vec\g)^{n_6}({\rm
sign} (\vec\a\times\vec\b\cdot\vec\g))^z,\; n_i=0,1,2,\dots,\;z=0,1 \}
\eqno(4.11)
$$

\ni could in principle serve as a basis for the gauge-invariant Hilbert
space, it is difficult to incorporate ${\rm
sign}(\vec\a\times\vec\b\cdot\vec\g)$ in integrations, since it is not
a smooth function on the classical group manifold. Another admissible
choice is

$$
\{\a_0^{n_1}\b_0^{n_2}\g_0^{n_3}
(\vec\a\cdot\vec\b)^{n_4} (\vec\a\cdot\vec\g)^{n_5}
(\vec\a\times\vec\b\cdot\vec\g)^{n_6},\; n_i=0,1,2,\dots\},
\eqno(4.12)
$$

\ni but this is not symmetric with respect to the three lattice
directions. A choice that solves both of these problems is

$$
\eqalign{
\{ e(n_1,n_2,n_3,n_4,n_5,n_6,z):=\a_0^{n_1} \b_0^{n_2} \g_0^{n_3}
(\vec\a\cdot\vec \b)^{n_4} (\vec\a\cdot\vec\g)^{n_5}&
(\vec\b\cdot\vec\g)^{n_6}
(\vec\a\times\vec\b\cdot\vec\g)^z,\cr
& n_i=0,1,2,\dots,\;z=0,1 \}. }\eqno(4.13)
$$

It contains all gauge-invariant information about the original Hilbert
space, without being overcomplete. There is a corresponding,
non-orthogonal basis for the square-integrable, holomorphic wave
functions, obtained by substituting as usual the real parameters in
(4.13) by their complex counterparts. Acting with the Hamiltonian
$\hat H^\C$ on such a state may produce terms that contain higher-order
powers $(\vec\a^\C\times\vec\b^\C\cdot\vec\g^\C)^n,\; n>1$,
which then have
to be re-expressed as a sum of (complex)
terms of the form (4.13) using the identity

$$
\eqalign{
(\vec  \a^\C&\times\vec\b^\C\cdot\vec\g^\C)^2=(1-(\vec\a^\C)^2)
(1-(\vec\b^\C)^2)(1-(\vec\g^\C)^2)-
(1-(\vec\a^\C)^2)(\vec\b^\C\cdot\vec\g^\C)^2\cr
&-(1-(\vec\b^\C)^2)
(\vec\a^\C\cdot\vec\g^\C)^2-
(1-(\vec\g^\C)^2)(\vec\a^\C\cdot\vec\b^\C)^2+
2(\vec\a^\C\cdot\vec\b^\C)
(\vec\a^\C\cdot\vec\g^\C)
(\vec\b^\C\cdot\vec\g^\C). }\eqno(4.14)
$$

This is an unambiguous prescription and leads to a Hamiltonian action
that maps states of the form $\sum
m(n_1,n_2,n_3,n_4,n_5,n_6,z)e(n_1,n_2,n_3,n_4,n_5, n_6,z)^\C$ into
themselves. However, as already mentioned, the  Hamiltonian is a
lengthy expression and solving the zero-eigenvalue problem is
certainly not a straightforward task.  This is probably just a
reflection of the non-triviality of quantum gravity. Still we
cannot rule out the existence of a basis for the gauge-invariant
Hilbert space that leads to a simplification of the eigenvalue
problem. However, this would presumably be an overcomplete basis,
which then leads to problems of a different kind. Firstly, one cannot
just set the coefficients of each ``basis" wave function to zero, and
secondly one has to eliminate spurious solutions by hand. By
contrast, an advantage of our formulation is that the general
condition on the wave function coefficients $m(\vec n)$ is known
explicitly, so that one has full control over all the relevant
physical parameters. This is important if one for instance decides to
introduce a cut-off in the Hilbert space, in order
to approximate the problem
of finding zero-eigenvectors by a finite-dimensional one.

\vskip2cm

\line{\ch 5 Conclusions\hfil}

We have described above a regularized version of non-perturbative
canonical quantum gravity on a cubic lattice with periodic boundary
conditions. The quantum Hamiltonian on the lattice acts
combinatorially on holomorphic wave functions labelled by
lattice loops. The lattice represents an entire diffeomorphism
equivalence class and the formalism is manifestly gauge- and
diffeomorphism-invariant at the kinematical level. With a specific
choice for the discretized Hamiltonian and a factor ordering in
the quantum theory, we are able to identify an
infinite-dimensional space of solutions to the discretized
Wheeler-DeWitt equation, which moreover have finite norm with respect
to a natural scalar product on the space of holomorphic
$SL(2,\C)$-functions. The solution space is labelled by global
Polyakov loops and their multiples, and corresponds to $\sim 3N^2$
physical degrees of freedom (compared to $\sim 6N^3$ before imposing
the Hamiltonian constraint).
Still, more research is needed to determine whether this
exhausts the solution space. For the example of the $1\times 1\times
1$-lattice, we illustrated how one may go about a systematic search
for more solutions. Although we have not yet been able to find any,
preliminary results suggest there may be an infinite set of solutions
beyond the Polyakov ones. However, even if that is the case, it may
still happen that they are not square-integrable.

Since our lattice regularization is rather different from the
point-split regularization used in the formal continuum approaches,
it is remarkable that our solution space is reminiscent
of the smooth, non-intersecting solutions of [6] (thus suggesting that
these in fact are not ``spurious").
We expect this feature to be fairly robust under a change of the
classical discretized Hamiltonian, because the existence of the
solutions depends only on the antisymmetric structure of $\e^{ijk}p_j
p_k$.  However, we have not found straightforward analogues of the
intersecting-loop solutions of [15,19]; this can be traced back to the
non-locality of the lattice Hamiltonian. The great advantage of our
regularization is the existence of a well-defined scalar product at
every stage. This will be crucial in all further investigations of
the solution space. For example, it would be interesting to understand
how a different factor ordering of $\hat H^\C$ changes our results.

In order to avoid confusion, it should be pointed out that we are
proceeding somewhat differently from the path outlined (for the
continuum theory) by Ashtekar and collaborators [20,13]. They propose
to solve the gauge and diffeomorphism constraints within a real
framework, based on the spin connection $\Gamma_a^i$, and then
go to a holomorphic representation to solve the Hamiltonian
constraint, whereas our formulation takes place entirely within the
complex formulation based on $A_a^i=\Gamma_a^i-i\, K_a^i$ (where
$K_a^i$ is related to the extrinsic curvature via $K_a^i=K_{ab}
E^{bi}$). This does not exclude that a close relation may emerge at the
level of dynamics.

Finally, as already mentioned in the introduction, one has to
face the question of the continuum limit (not to be confused with a
weak-field limit) of the regularized  theory. The diffeomorphism
invariance of general relativity makes this a fundamentally different
issue from that in lattice gauge theory. If one wants to avoid
bringing back in an ultraviolet cutoff $a$, the only free
parameter is the lattice size $N$, and one would expect that
in the limit for growing $N$ a prospective continuum theory is
approximated ever better. Since (at least part of) the solution
space is known for every finite $N$ (and the $N$-dependence of our
construction is rather explicit), one might investigate the limit as
$N\rightarrow\infty$ of these spaces directly. This will
probably become more meaningful once observables and possibly matter
fields have been included, so that one can study their spectral
properties as a function of $N$. Our construction suggests that in
such a ``continuum" limit some fundamental discrete structure is
retained, although (as pointed out in [21]) this does not necessarily
preclude the appearance of divergences.

\vfill\eject

\line{\ch References\hfil}

\item{[1]} Jurkiewicz, J.: Simplicial gravity and random surfaces,
  {\it Nucl. Phys.} B {\it Proc. Suppl.} 30 (1993) 108-21

\item{[2]} Renteln, P.: Some results of SU(2) spinorial lattice
  gravity, {\it Class. Quant. Grav.} 7 (1990) 493-502

\item{[3]} Piran, T. and Williams R.: Three-plus-one formulation of
  Regge calculus, {\it Phys. Rev.} D33 (1986) 1622-33;
  Friedman, J.L. and Jack, I.: 3+1 Regge calculus with
  conserved momentum and Hamiltonian constraints, {\it J. Math. Phys.}
  27 (1986) 2973-86;
  Renteln, P. and Smolin, L.: A lattice approach to spinorial
  quantum gravity, {\it Class. Quant. Grav.} 6 (1989) 275-94;
  Bostr\"om, O., Miller, M. and Smolin, L.: A new
  discretization of classical and quantum general relativity, {\it
  preprint} G\"oteborg U. ITP 94-5 and Syracuse U. SU-GP-93-4-1

\item{[4]} Ashtekar, A.: New variables for classical and quantum
  gravity, {\it Phys. Rev. Lett} 57 (1986) 2244-7; A new
  Hamiltonian formulation of general relativity, {\it Phys.
  Rev.} D36 (1987) 1587-1603

\item{[5]} Ashtekar, A.: {\it Lectures on non-perturbative canonical
  gravity}, World Scientific, Singapore, 1991

\item{[6]} Rovelli, C. and Smolin, L.: Loop space representation of
  quantum general relativity, {\it Nucl. Phys.} B331 (1990) 80-152

\item{[7]} Br\"ugmann, B.: Loop representations, in {\it Canonical
  gravity: from classical to quantum}, ed. J. Ehlers and
  H. Friedrich, Lecture Notes in Physics 434, Springer, Berlin,
  1994

\item{[8]} Ashtekar, A. and Lewandowski, J.: Projective techniques
  and functional integration for gauge theories, to appear in:
  {\it J. Math. Phys.} special
  issue on {\it Functional Integration}, ed. C. DeWitt-Morette,
  e-Print Archive: gr-qc 9411046

\item{[9]} Kucha\v r, K.V.: Covariant factor ordering of
  gauge systems, {\it Phys. Rev.} D34 (1986) 3044-57;
  see also C.J. Isham: Quantum gravity, in {\it General Relativity and
  Gravitation}, Proceedings of the 11th International Conference,
  ed. M.A.H. MacCallum, Cambridge University Press, Cambridge, 1987

\item{[10]} McLerran, L.D. and Svetitsky, B.: A Monte Carlo study
  of $SU(2)$ Yang-Mills theory at finite temperature, {\it Phys.
  Lett.} 98B (1981) 195-8; Kuti, J., Pol\'onyi, J. and
  Szlach\'anyi, K.: Monte Carlo study of $SU(2)$ gauge theory at
  finite temperature, {\it Phys. Lett.} 98B (1981) 199-204

\item{[11]} Kogut, J. and Susskind, L.: Hamiltonian formulation of
  Wilson's lattice gauge theories, {\it Phys. Rev.} D11 (1975)
  395-408; Kogut, J.B.: The lattice gauge theory approach
  to quantum chromodynamics, {\it Rev. Mod. Phys.} 55 (1983) 775-836

\item{[12]} Hall, B.C.: The Segal-Bargmann coherent state transform
  for compact Lie groups, {\it Journ. Funct. Anal.} 122 (1994) 103-51

\item{[13]} Ashtekar, A., Lewandowski, L., Marolf, D., Mour\~ao, J.
  and Thiemann, T.: Coherent state transform for spaces of connections,
  {\it preprint} Penn State U., Dec 1994, e-Print Archive: gr-qc
  9412014

\item{[14]} Br\"ugmann, B., Gambini, R. and Pullin, J.: Jones
  polynomials for intersecting knots as physical states of quantum
  gravity, {\it Nucl. Phys.} B385 (1992) 587-603

\item{[15]} Jacobson, T. and Smolin, L.: Nonperturbative quantum
  geometries, {\it Nucl. Phys.} B299 (1988) 295-345

\item{[16]} Creutz, M.: {\it Quarks, gluons and lattices}, Cambridge
  University Press, Cambridge, 1983

\item{[17]} Loll, R.: Independent SU(2)-loop variables and the reduced
  configuration space of SU(2)-lattice gauge theory, {\it Nucl. Phys.}
  B368 (1992) 121-42; Yang-Mills theory without Mandelstam
  constraints, {\it Nucl. Phys.} B400 (1993) 126-44

\item{[18]} Watson, N.J.: Gauge invariant variables and Mandelstam
  constraints in $SU(2)$ gauge theory, {\it preprint} Marseille
  CPT-94-P-3065, Aug 1994, e-Print Archive: hep-th 9408174

\item{[19]} Husain, V.: Intersecting loop solutions of the
  Hamiltonian constraint of quantum general relativity, {\it Nucl.
  Phys} B313 (1989) 711-24; Br\"ugmann, B. and Pullin, J.:
  Intersecting N loop solutions of the Hamiltonian constraint of
  quantum gravity, {\it Nucl. Phys.} B363 (1991) 221-44

\item{[20]} Ashtekar, A.: Recent mathematical developments in quantum
  general relativity, to appear in {\it Proceedings of the VIIth
  Marcel Grossmann Conference}, ed. R. Ruffini and M. Keiser,
  World Scientific, Singapore, 1995, e-Print Archive: gr-qc
  9411055

\item{[21]} Rovelli, C. and Smolin, L.: Discreteness of area and
  volume in quantum gravity, {\it preprint} Pittsburgh U. and
  Penn State U., Nov 1994, e-Print Archive: gr-qc 9411005

\end